# The Foundations of Computational Management:

## A Systematic Approach to Task Automation for the Integration of Artificial Intelligence into Existing Workflows

**Tamen Jadad-Garcia**
Vivenxia Group LLC, Beverly Hills, California, USA
tjadadgarcia@vivenxia.com

**Alejandro R. Jadad, MD DPhil LLC**
Founder, Centre for Digital Therapeutics, Toronto, Canada; Research Professor (Adjunct), Department of Population and Public Health, University of Southern California, Los Angeles, USA
aj_492@usc.edu or ajadad@gmail.com

## Abstract

Driven by the rapid ascent of artificial intelligence (AI), organizations find themselves at the epicenter of a seismic shift, facing a crucial question: How can AI be successfully integrated into existing operations? To help answer this question, manage expectations and mitigate frustration, this article introduces Computational Management, a systematic approach to task automation for enhancing the ability of organizations to harness AI's potential within existing workflows. Computational Management acts as a bridge between the strategic insights of management science with the analytical rigor of computational thinking. The article offers three easy step-by-step procedures to begin the process of implementing AI within a workflow. Such procedures focus on task (re)formulation, on the assessment of the automation potential of tasks, on the completion of task specification templates for AI selection and adaptation. Included in the article there are manual and automated methods, with prompt suggestions for publicly available LLMs, to complete these three procedures.

The first procedure, task (re)formulation, focuses on how to break down work activities into their most basic units, so that they can be completed independently by one agent, involve a single well-defined action, and produce a distinct identifiable outcome. The second procedure allows the assessment of the granular task and its suitability for automation, using the 'Task Automation Index'. This is a practical tool that helps rank the tasks based on the extent to which they have standardized input, well-defined rules, repetitiveness, data dependency, and objective outputs. The third procedure focuses on how to complete a 'task specification template' which details explicit information on the 16 critical components of tasks with high suitability for automation. This template can be used as a checklist to identify, select or adapt the most suitable AI solution for seamless integration into existing workflows.

Computational Management thus provides both a roadmap and a new toolkit for navigating the transformative impacts of AI, contributing to the creation of a future where humans and AI can thrive together, enhancing organizational efficiency and innovation.

# Introduction

Organizations find themselves at the epicenter of a seismic shift driven by the rapid ascent of artificial intelligence (AI). The potential to harness AI for enhanced productivity and efficiency is vast, yet a crucial question lingers for decision-makers: How can we successfully integrate AI into our existing operations?.

Fulfilling the transformative potential of AI within organizational settings requires the navigation of complex multifaceted challenges. These span the entire operational spectrum of organizations—from strategic planning at the executive level to the day-to-day activities of any enterprise—and they extend through the entire AI development lifecycle—from initial concept through development, deployment, and ongoing management. To overcome these challenges, it is essential to commit to a fundamental shift in perspective. We must move away from a traditional job-centric lens, which views roles in terms of broad job descriptions, titles and skills, to a granular focus on tasks, the most basic and actionable units of work, in order to integrate AI successfully (1).

This shift is pivotal for several reasons. First, a job-centric view tends to underplay the specific and fundamental activities that take place within an organization—a form of 'oblivious to the obvious bias' (2). Jobs are often described in broad strokes that encompass a range of roles and tasks, some of which may be amenable to automation with AI, while others may not. This high-level vantage point sets the scene for unrealistic expectations about what AI can and cannot do, and can lead to wasted investments.

In contrast, a task-oriented perspective breaks down jobs into their basic building blocks of activities, offering a clearer view of the work that makes up existing workflows and provides a more actionable foundation for integrating AI (3). For example, while a job may involve a mix of analytical, creative, and interpersonal activities, a task-focused analysis might reveal that the analytical components—such as data processing or pattern recognition—are particularly suited to augmentation or automation with AI.

Focusing on tasks rather than skills, roles or jobs makes it possible to have a more agile and precise approach to AI integration. It enables organizations to pilot AI solutions on a task-by-task basis, iteratively scaling successful implementations and adjusting or abandoning those that do not meet expectations. A task-centric approach also facilitates the identification of suitable AI applications and enhances the adaptability of the organization to technological advancements, ensuring that AI integration efforts are both strategic and sustainable. By making this fundamental shift in perspective, organizations can more effectively navigate the complexities of the AI landscape, tailor AI solutions to their specific needs, and drive meaningful improvements in productivity and efficiency.

Given the value to be unlocked with this perspective shift, it is essential to be intentional with how tasks are formulated. Task descriptions optimized for human interpretation limit AI's potential. For instance, a task for a sales manager position (4), such as "direct and coordinate activities involving sales of manufactured products, or services" poses significant challenges for effective automation. Tasks that are broad, and that include multiple actions or multiple results, hinder the ability for AI to meet expectations. Furthermore, this bundling of activities makes it very difficult to identify and implement effective AI tools capable of being integrated in existing workflows or meeting the specific needs of an organization. It also creates the conditions for frustration among decision-makers and managers and sets-up new technology to fail. This requires a reimagining of tasks to make them actionable and suitable for both humans and AI (5).

Once the clarity of tasks within an organization is established, delineating the specific actions and outcomes expected, there is another layer of complexity inherent to task execution. This complexity stems from the multifaceted nature of tasks themselves, which comprise several critical components essential for their successful completion. These components include the actions that need to be performed, the sequential steps that guide these actions, the capabilities or skills required to execute them, the materials necessary to carry out the task, a clear description of what the final product should look like, and the criteria that determine whether the task has been satisfactorily completed. Breaking down tasks into these granular elements serves a dual purpose: it streamlines the communication of expectations, ensuring that everyone involved has a clear understanding of what needs to be done, and it systematically equips agents—those responsible for carrying out the tasks—with the roadmap and resources needed for success.

In this context, therefore, the term 'agent' refers to any individual or entity capable of executing tasks. Within the diverse ecosystem of an organization, agents encompass a broad spectrum of roles, including front-line operators, managers, executives and members of the board. This inclusive definition acknowledges the contribution of every organizational member in the collective effort to achieve goals and deliver outcomes. However, a significant hurdle to leveraging AI to enhance task execution is the lack of specialized tools designed to deconstruct tasks into these detailed components in a systematic, replicable and scalable way, which take into account the requirements of humans and machines to complete tasks effectively and efficiently. The absence of such tools not only impedes the effective integration of AI into existing workflows but also limits the potential for AI to augment human efforts and drive efficiency.

Given that the method of breaking down tasks is so crucial, the challenge extends beyond merely identifying tasks suitable for AI integration. It also involves developing and implementing a systematic approach to dissect tasks into their constituent parts, with the intention of making them more accessible for AI solutions to tackle. This detailed task breakdown is essential for several reasons: it offers a systematic and structured way of dissecting tasks that is useful for both AI tools and human agents, it facilitates a deeper understanding of a workflow, it enables the precise mapping of AI capabilities to task requirements, and it prepares the groundwork so that all prerequisites for task automation and completion can be met. By addressing this gap, organizations can more effectively harness AI's potential within existing workflows, not only to automate repetitive or mundane tasks but also to augment human capabilities in executing more complex tasks.

Efforts to integrate AI within organizational settings could also benefit from the availability of informative and actionable tools that can be applied on the ground, by managers and executives, to assess automation potential of tasks. Such a strategic assessment tool would not only simplify decision-making but also unlock operational efficiencies by identifying prime candidates for automation that could have otherwise been overlooked. In this way, it is possible to better allocate human and AI resources, where each can contribute most to achieving operational excellence and drive growth. At the same time, such tools could empower AI developers to tailor their solutions to achieve a greater fit with organizational efforts to automate tasks, optimize workflows and boost their productivity.

The rapid advancements in AI and robotics are bringing about an unprecedented convergence between management and computer science, heralding a new era of hyper-productivity and growth. This dynamic fusion is inviting organizations across all industries to reimagine their operations, embedding cutting-edge technological insights at the core of business strategies. This evolution signifies a shift towards more intelligent, agile, and technology-driven business models, reshaping the foundational aspects of how organizations operate and thrive in the digital age.

In an effort to help address all of these challenges and opportunities presented by the era of AI, in this paper we introduce an approach called Computational Management, a systematic approach to task automation for enhancing the ability of organizations to harness AI's potential within existing workflows. This approach acts as a bridge between the strategic insights of management science with the analytical rigor of computational thinking. By leveraging structured, transparent, and adaptable procedures, Computational Management aims to make it easier and more convenient to adopt AI and benefit from this era rich in technological advancements.

In this article, we delve into the components of Computational Management and offer practical step-by-step procedures—manual and automated (using LLMs)—to implement AI into existing operations. We present methods to: (re)formulate tasks, evaluate automatability, and complete a standardized template that helps to select and adapt AI offerings. Computational Management intends to become a pivotal resource for navigating the complex landscape of AI integration within organizations, and for optimizing human-AI collaboration.

## Methods and Procedure

Given that Computational Management is a new process, creating a framework for it requires drawing on the basic concepts from precursor frameworks in computational thinking and management science. This enables the identification of common solid ground to build what is needed by organizations to successfully integrate AI tools into their existing workflows.

A conceptual framework is a network of structured and interconnected ideas and other imaginary entities (collectively known as 'concepts'), which can be expressed using words or images in order to make a phenomenon understandable (6).

To be eligible for inclusion, a precursor framework had to include: a description of the sources used to find the supporting evidence used to identify the component ideas or concepts of computational thinking or management science; a narrative description or a graphical representation of the network of such ideas, in relation to the phenomenon of interest; and at least one concept related to the process of enabling human or AI agents to complete tasks in organizations. Our decisions were guided by the yield of systematic searches of the scholarly and gray literature using successive Boolean strategies on Google Scholar, and on the large language models (LLMs) Bard/Gemini, GPT-4/Bing/Copilot and Claude, that were run initially on October 25, 2023, followed by weekly updates up to February 1, 2024 (Appendix A).

The initial search strategy was the broadest. Using the terms intitle: *"computational management"*, Google Scholar yielded 54 hits, which were reviewed by both of us, separately and jointly, confirming the lack of a conceptual framework focused on it. We found two books (7,8), a book series (9), a journal (10) and a doctoral thesis (11) with 'computational management' in their titles, of which none provided a conceptual framework. The only conceptualization or definition of Computational Management that applied directly to the workplace appeared at the beginning of the aforementioned book. Such a definition was circular and lacked operationalizability, as it read:

*"Since most of the business processes have computational aspects hence, they can be represented mathematically so as to make it easy to incorporate intelligence based approaches to the systems. Thus, this management of the computational aspects of various management problems that lead to the conceiving of the term Computational Management"* (12)

The search strategy string: *intitle:"conceptual framework" AND intitle:"computational thinking"* identified 15 separate conceptual frameworks, none of which focused on the management profession, on management sciences, on the organizational setting, or on the completion of tasks, or the adoption of AI. Nevertheless, one of the frameworks provided an operationalizable approach for defining computational thinking (13). This framework was built with 25 concepts identified in the literature and three new concepts proposed by the authors, who displayed them as a generic network illustrating, in a novel way, how agents (human or computer) complete tasks after they receive, gather and use information (inputs) to guide their actions (sequential algorithmic steps) with the purpose of generating a result or outcome (output).

Computational thinking has been conceptualized as the systematic process of formulating goals with enough clarity that it becomes possible to tell an information-processing agent—human or computer—how to achieve them (14,15). It relies on the application of key principles and techniques from computer science, mathematics and engineering to tackle complex, multi-faceted problems across diverse domains.

The process of computational thinking entails breaking down real-world ambiguous challenges into explicitly defined tasks that agents can analyze to generate solutions efficiently. The rationale behind this process is that "learning to think like a computer scientist would be a benefit for everyone, in whatever profession …[by giving the professionals] a lens and a set of categories for understanding the algorithmic fabric of today's world… and for 'making the abstract concrete.'"(16).

Computational thinking rests on the following pillars to tackle problems:

- Abstraction: the formal extraction of the essential details from messy real-world conditions that impact the problems, in order to encode only the most salient stimulating factors and measurable outcomes related to possible solution paths.

- Decomposition: the rigorous unpacking of the overall challenge into explicit logical components and sub-questions in such a way that computational agents could systematically operate upon them, within a more delimited and tractable problem space.

- Automation: the codification of a series of deterministic rules to transform well-defined input states into desired output solutions in a methodical step-by-step sequence. Defining automated pathways illuminates options through systematic simulation.

- Evaluation: the use of quantitative or qualitative methodologies to assess results and judge the level of improvement achieved.

- Iteration: the implementation of cycles of sequences of steps based on failures and successes, to refine assumptions, rerun solutions, and test changes aimed at increasing the understanding of the problem and its solution.

Another search strategy was designed and run to identify precursor conceptual frameworks focused on management science (Appendix A). It yielded 120 potentially eligible documents, of which one met the selection criteria. This focused on the Occupational Information Network (O*NET), a framework developed and maintained by the US Department of Labor to support the standardization and digitization of information about all occupations in the nation's labor market. The framework—based on scientific management principles, as well as human relations, socio-technical and sociological constructs—supports the standardization and guides the creation of an interactive structured digital platform that can be used to describe more than 1,000 occupations (17). This framework and its platform are also used as the basis for writing job descriptions, and for guiding job design and reskilling efforts by businesses across the country and around the world (18).

The Content Model consists of six major domains, which are further divided into sub-domains and elements. The domains are:

- Worker Characteristics: These are the enduring qualities of individuals that may influence how they approach tasks and how they acquire work-relevant knowledge and skills. Examples of elements in this domain are abilities, occupational interests, work values, and work styles.

- Worker Requirements: These are the attributes that workers must have, or must acquire, to perform tasks successfully. Examples of elements in this domain are skills, knowledge, and experience.

- Occupational Requirements: These are the characteristics of work that describe what various occupations (clusters of jobs including similar tasks) require of workers. Examples of elements in this domain are work activities, organizational context, and work context.

- Experience Requirements: These relate to previous work activities and explicitly linked to certain educational and training programs. Examples of elements in this domain are experience and training, basic entry skills, and licensing.

- Occupation-Specific Information: These are the descriptors that are unique to each occupation. Examples of elements in this domain are title, description, alternate titles, and technology skills and tools. Here is where more than 20,000 tasks distributed across all of the included occupations are listed.

- Workforce Characteristics: These are the descriptors that are relevant to a group or class of occupations. Examples of elements in this domain are labor market information and occupational outlook.

O*NET is used by researchers to explore the implications of AI in organizations, including those related to job augmentation (19), exposure (20) or automation (21,22), as well as the coevolution of AI and tasks in the workplace (23). The focus of the latter on tasks reflects concerns by researchers about the risk of bias associated with focusing narrowly on workers' skill-levels, which misses the full picture of how technology changes employment. Skills describe what individual capabilities people have. However, technologies get directly built into the tasks, which are the smallest unit of work activity that are bundled into jobs (24). For example, a public relations and communications job involves tasks like generating copy for products or services, preparing reports and interacting with clients. AI can impact these kinds of tasks, regardless of the level of skill required.

This task-focused approach is crucial, as it addresses a key issue that is constantly lurking under the surface of deliberations about the readiness of the workplace for the adoption of AI: Once it becomes technically feasible for a human or an AI agent to perform a task, the one with the lowest financial cost will be chosen (25).

Given their prominent role in the preparation of the workplace for AI, we ran an additional search strategy (Appendix A), which identified two additional frameworks that provided in-depth analyses of tasks (26, 27), both from a management science perspective.

One of these frameworks (26) viewed tasks as the basic unit of work-related activity, that result from breaking down complex goals or objectives into smaller, manageable and actionable units, so that they could be executed more efficiently by individual or teams of agents charged with achieving a specific goal or outcome. Within this framework, tasks share three essential components:

- Cues: The signals, data, or prompts that provide contextual guidance for agents to guide the initiation of task-related acts. They are structured using text, numbers, images or audiovisual files.

- Acts: The steps and patterns of behaviors with identifiable purpose or direction taken in response to cues.

- Results: The end products or outcomes of the acts.

This approach from management science has many similarities with how tasks are viewed in computer science. In both cases, tasks are identified by breaking down complex problems into smaller, more manageable units, so that they specify the actions that need to be performed to achieve a desired result. Instead of dividing tasks into cues, acts and results as in the management science, tasks are usually divided into inputs, rules or algorithms, and outputs in computer science. Although the language of computer science is often considered as more rigorous, precise and rigid than that of management, major technological breakthroughs in AI and robotics are demanding greater consistency between them. Given that they are acknowledged as the smallest units of activity within management and computer science, tasks are a logical meeting point for efforts to prepare organizations for AI. This is why we decided to place them as the core component of the Computational Management framework.

As an initial step, we generated a list with the 18 generic task descriptions related to 'Corporate Communications Specialist' (28) using the Job Description Writer of O*NET (Appendix B). We then examined each of the descriptions. Just a cursory review of each of the descriptions led us to conclude that they lacked enough detail to enable agents, regardless of whether they are human or AI, and especially the latter, to complete such tasks. An example that illustrates this is the task formulated as: *"Plan or direct development or communication of programs to maintain favorable public or stockholder perceptions of an organization's accomplishments, agenda, or environmental responsibility."*

Given that O*NET's current task descriptions, geared for human interpretation, limit AI's potential in organizations—and that it would be impossible to use them to match the specific needs of organizations with tools that enable automation—we decided to build a modular system with three generic machine-friendly stand-alone procedures. These are designed to set up AI tools and organizations for success by reducing the risk of misalignment due to lack of clarity in language and expectations.

The first procedure focuses on task (re)formulation; the second on assessing and ranking the tasks based on their ability to be automated; and the third, on filling out the task specification template, in order to identify or adapt the best AI tools to successfully integrate them into existing workflows. We will begin with a description of the fundamentals behind each of these procedures, and then share the step-by-guide as to how to put it all together (either manually or using LLMs).

**Task (re)formulation**

This procedure considers tasks as the most basic units of work. The purpose of (re)formulation is to enhance clarity for agents, particularly those driven by AI, as well as to boost organizational efficiency and accountability through optimized, more transparent workflows, ensuring that each task is sufficiently discrete to be independently manageable and measurable, and executable by an agent.

Within this context, a task would be regarded as comprising a basic unit of work when it meets ALL of the following criteria:

- Completable by a single agent independently (with or without consulting or others).
- A single well-defined action represented by a single verb.
- A distinct identifiable outcome, either material or informational, which serves as a clear indicator of completion.

For example, "Draft the schedule for trade shows this year" is a task, whereas "Coordinate or participate in promotional activities or trade shows, working with developers, advertisers, or production managers, to market products or services" is not, as it meets none of the criteria.

**Assessing ease of automation**

Once tasks are broken down into the most basic units of work, it is possible to start evaluating their suitability for automation. This could be achieved by examining the extent to which the task fulfills the following conditions, which have been selected because of their relevance across various industries, their applicability to both physical and mental tasks, and their direct relevance to the implementation of AI into existing operational workflows:

- Standardized input: This component assesses whether the agent receives information that is consistent, structured, and presented in a format that the agent can accurately understand and process. An input is consistent when it has a predefined format specifying the order and naming of fields and data types (e.g. text, numbers, dates). This is key for efficient and error-free processing by automated systems, especially machine learning algorithms. Inconsistent or varying inputs can lead to inaccuracies in AI processing, as the system may struggle to recognize patterns, parse information, or draw correct conclusions. Standardization reduces the need for complex preprocessing and error handling, making the AI more efficient and reliable.

- Well-defined rules: This component focuses on whether the steps to complete a task are unambiguous and organized in a logical order. Well-defined rules are essential for determining if they can be codified for AI to follow with high accuracy and minimal human intervention.

- Repetitive: This component looks at whether the task is performed frequently, involving similar processes or actions each time. High repetitiveness makes a task more suitable for automation, as it allows AI systems to identify patterns, learn and guide its actions, in order to optimize their consistency and efficiency, while reducing the need for human intervention.

- Data-dependent: This component focuses on data availability and data requirements to complete a task. Tasks that rely heavily on numerical data or other quantifiable information are more amenable to automation. This also enhances the ability of AI systems to learn, make predictions, and improve over time, leading to more efficient and effective decision-making.

- Objective output: This component refers to whether the task's outcome can be verified. Tasks with clear metrics or benchmarks for their outputs, which can be objectively verifiable, are more suitable for automation because they allow for clear performance assessment and validation. They also enable AI systems to undergo fine-tuning over time.

We have generated a 'Task Automation Index' that can be used to systematically evaluate tasks against these criteria, so that organizations can better understand the potential for automating various aspects of their operations, paving the way for more effective and seamless AI integration (see Appendix C).

After ranking tasks in terms of their suitability for automation, at this higher level, the task specification template could be used to help identify, select or adapt AI solutions that can best be integrated into existing workflows.

## Task specification template

Given that tasks are regarded—consistently in both the computer sciences and management science literature—as the smallest unit of activity, they could be viewed as analogous to atoms in the physical world. Following this analogy, the component parts of a task would be equivalent to subatomic particles, like electrons, protons and neutrons. When drawing parallels with this idea, an assemblage of cues, acts and products gives rise to tasks.

By continuing to zoom out from cues to tasks, it is possible to see how Computational Management can have an impact at scale; multiple tasks, make up jobs, which then allow the creation of teams; groups of teams form organizational units, which together turn into organizations. Organizations form industries, and these are ultimately clustered into sectors.

To ensure that an agent can complete a task, it is important to have a complete task specification template which captures the most fundamental information about what is required. Given the almost complete lack of background evidence, we generated, by consensus, the following list of 16 component parts for the task specification template, on the basis of our own judgment and the limited literature available:

- Unique ID: This is a value or code that identifies a specific completed task specification template, using numeric or alphanumeric strings.

- Agent: The one responsible for task completion.

- Deadline: This should be an unambiguous target calendar date, specified in MM/DD/YYYY numerics, representing the date before which the task should be completed.

- Action: The act or behavior that is required to perform the task. It should be simple, concrete, and achievable. It should also be aligned with the task name, with its focus on one verb.

- Steps: These are the instructions for sequential smaller activities that require orchestration among people and tools, and that are mapped to generate the main output.

- Frequency: Expected regularity of task activation based on the recurrence of the same task.

- Materials: Checklist of raw work elements necessary for commencing and completing the task.

- Result: A single noun with clear descriptors qualifying the minimum acceptable work result following the execution of all steps.

- Capabilities: Enumeration of skills, tools or traits required for the fulfillment of the task.

- Completion criteria: Product specifications that determine whether an output meets minimum quality standards to declare the task completed.

- Completion instructions: Protocols guiding handoff of task products to relevant downstream or upstream stakeholders.

- Report summary: A brief overview or summary, including the task name, the task product, and the task status.

- Report language: The report language is the language or dialect that is used to communicate about the task.

- Report medium: The format or type of the task report, including text, audio, video or image.

- Delivery mode: The method or channel that is used to deliver the task report.

- Status options: The possible states or conditions of the task after execution of the rules. They can be completed, in progress, pending, canceled.

We then decided to use LLMs to ensure that this was comprehensive and included all of the relevant parts. Thus, we primed Bard/Gemini, Claude and GPT-4/Bing/Copilot with the conceptualizations of Computational Management, tasks and their constituent components, and the rationale for the exploration of the task specification template and its structure.

We then prompted the LLMs with a request to examine the above list and to suggest additional parts, if appropriate. Consistently, the LLMs considered the list to be comprehensive. In addition, each of the LLMs suggested additional items for our consideration, all of which were excluded either because they were regarded as optional, potentially distracting or unnecessarily time-consuming for the agents, or too dependent on the specific characteristics of the relevant task.

The task specification template has been designed to help evaluate or adapt AI options and integrate them into existing workflows.

## Putting it all together

Anyone interested in implementing AI into their current operations could benefit from following a four-part process: First, (re)formulate a task into the most basic units of work; second, identify the easiest tasks to automate; third, generate the characteristics and functionalities required for an agent using the task specification template; and fourth, seek the best option given the context. To complete these parts, we offer a manual procedure and an automated approach that uses the power of available LLMs below.

**Manual method**

*Part 1: Task (re)formulation*

- Step 1: Read the original description of the task and determine whether it meets the criteria for a basic unit of work (being completable independently by a single agent, having a well-defined action represented by a single verb, and possessing a distinct, identifiable outcome), without needing further breakdown.

    *For example, the task description "Coordinate or participate in promotional activities or trade shows, working with developers, advertisers, or production managers, to market products or services"*

- Step 2:
    - Step 2.1: If the task description does not meet the criteria for the most basic unit of work, break it down into smaller and simpler components, based on the criteria. Use bullet points to list the basic units of work for each description, making sure that each of them meets the task criteria.

    *In the example, the task description "Coordinate or participate in promotional activities or trade shows, working with developers, advertisers, or production*

*managers, to market products or services" can be broken down into the following units of work (the focus will be on participating in promotional activities or trade shows, rather than coordinating them):*

- Identify potential trade shows or promotional activities relevant to the product or service.
- Gather product features and updates from developers.
- Develop marketing strategies with advertisers for promotional activities.
- Confirm product availability with production managers.
- Organize presentation materials for events with production managers.
- Book event spaces for trade shows or promotional activities.
- Arrange travel for participants.
- Arrange accommodation for participants.
- Schedule shipping of products and materials to the event location.
- Commission booth or display area design.
- Approve the booth or display area design.
- Set up booth or display area at trade shows or promotional activities.
- Demonstrate products or services to attendees.
- Collect contact information from potential clients or partners.
- Oversee budget adherence.
- Oversee deadline adherence.

*Each of these units of work has one verb and one output, and can be completed by one agent, independently, without needing further breakdown.*

- ○ Step 2.2: If the task description already meets the criteria for the most basic unit of work, then move to part 2.

*Part 2: Assessing and ranking the ease of task automation*

In this part of the process the aim is to use the Task Automation Index to generate a total score representing the extent to which a task reflects five conditions for automation. Each condition has a set of statements about a task with a designated amount of points for each (see Appendix C).

- Step 1: Create a table with seven columns and the amount of rows that represents the number of tasks that were generated in Part 1, plus an additional row for the headers. In the top row of the table, label the columns with the following headings from left to right: "Task", "Standardized input", "Well-defined rules", "Repetitive", "Data-dependent", "Verifiable or measurable output" and "Total score". Then fill out the first column on the left, under the "Task" heading, with the list of tasks generated in Part 1.

- Step 2: Beginning from the top of the list of tasks generated in Part 1, select the statement that best describes the the extent to which the task reflects the specific condition for automation, and write down the number of points under the corresponding heading:

*Standardized input:*

- 0 points: Input format is not predefined and data types are completely inconsistent.
- 1 point: Input format and data types are mostly inconsistent.
- 2 points: Input format and data types are somewhat standardized.
- 3 points: Input format is mostly predefined and data types are consistent with rare exceptions.
- 4 points: Input format is predefined specifying the order and naming of fields with consistent data types.

*Well-defined rules:*

- 0 points: There are no rules to complete the task.
- 1 point: The rules are ambiguous.
- 2 points: Rules are explicit for most scenarios, with areas of ambiguity.
- 3 points: Rules are explicit, with very few ambiguous aspects, and are mostly organized in a logical order.
- 4 points: Rules are explicitly defined and organized in a logical order.

*Repetitive:*

- 0 points: Task is unique or never repeated.
- 1 point: Task is occasionally repeated.
- 2 points: Task is often repeated.
- 3 points: Task is frequently repeated.
- 4 points: Task is always repetitive.

*Data-dependent:*

- 0 points: Data not necessary for task completion.
- 1 point: Data somewhat required or unavailable for task completion.
- 2 points: Data required and somewhat available for task completion.
- 3 points: Data required and mostly available for task completion.
- 4 points: Data required and fully available for task completion.

*Objective output:*

- 0 points: Output does not have a metric or benchmark, or cannot be verified
- 1 point: Output has limited metrics or benchmarks, or verification is mainly subjective
- 2 points: Output has clear metrics or benchmarks, but verification is mainly subjective.
- 3 points: Output has clear metrics or benchmarks and is mostly verifiable objectively.
- 4 points: Output has a clear metric or benchmark and is completely verifiable objectively.

- Step 3: Add up the total points in each row and write the result for each in the last column, under the heading "Total score".

- Step 4: In a list, or in the same table, rearrange the tasks in order of most automatable (those with the higher score) to least (those with the lower scores).

- Step 5: Interpret the results using the following suggested guidelines, included The Task Automation Index:

    16-20: Highly Suitable for Automation

    Tasks scoring in this range are excellent candidates for automation, indicating a high degree of standardization, clear rules, repetitiveness, data-dependency, and measurable outputs.

    12-15: Suitable for Automation

    These tasks are good candidates for automation, showing strong potential with some areas that might benefit from minor adjustments or refinements.

    8-11: Moderately Suitable for Automation

    Tasks within this score range are possible to automate but might require significant modifications to enhance their automation potential or might be more complex to automate effectively.

    4-7: Limited (Low?) Suitability for Automation

    Such tasks have limited automation potential due to significant variability, ambiguity, low repetition, or subjective outputs. Automation may be feasible for parts of the task or in a limited capacity.

    0-3: Not Suitable for Automation

    Tasks scoring in this lowest range are currently not viable for automation with existing technologies, likely requiring a high degree of human judgment, creativity, or adaptability.

*Part 3: Task specification template completion*
- Step 1: Select a task (in its most basic formulation), for example: *"Confirm product availability with production managers"*. It is recommended that you choose one that is considered to be easier to automate.

- Step 2: Review the list of task specification template parts and identify those that are relevant and applicable to the task. These are the components that provide detailed information about the task, such as the agent, the deadline, the action, the steps, the frequency, the materials, the result, the capabilities, the completion criteria, the completion instructions, the report summary, the report language, the report medium, the delivery mode, and the status options (refer to the list under the heading "Task specification template" above).

- Step 3: Create a list with the relevant parts formatting it as a form, table, or bulleted text, leaving space for adding task specific details to each item.

- Step 4: Fill out your structure with the appropriate information. For example, use:

  - Unique ID: Numeric or alphanumeric strings to assign to your completed task specification template, such as *"CPAPM-20240201"*.

  - Agent: A noun or a noun phrase to identify the agent responsible for the task completion, such as *"Name (First and Last for a human agent)]"* or *"Name (Product, version and company for a digital agent)"*, or *"TBD"* in the event that an agent has not yet been assigned.

  - Deadline: An unambiguous target calendar date, specified in MM/DD/YYYY numerics, to indicate the deadline for the task, such as "04/15/2024".

  - Action: Confirm product availability

  - Steps: A numbered list of simple, concrete, and achievable instructions to outline the sequential smaller activities that are required to generate the main output, such as:
    1. Identify product and quantity: Specify the product name, model number, and desired quantity.
    2. Select relevant production manager(s): Choose the production manager(s) responsible for the specific product line.
    3. Initiate communication: Contact the production manager(s) through preferred method(s) (e.g., email, phone call, instant message).
    4. Inquire about availability: Clearly ask about the availability of the desired product quantity within the specified timeframe.
    5. Receive confirmation: Obtain a clear confirmation of availability or an estimated delivery date if unavailable.
    6. Document confirmation: Record the confirmation details, including date, production manager contacted, and confirmed availability/delivery date.

  - Frequency: A word or a phrase to indicate the expected regularity of task activation based on the recurrence of the same task, such as *"Weekly or as required by inventory cycles and production schedules"*.

  - Materials: A bulleted list of raw work elements necessary for commencing and completing the task, such as:
    - Order details document
    - Access to the inventory management system.
    - Current production schedules.
    - Contact information for production managers.
    - Communication platform (e.g., email, phone)

- Documentation tool (e.g., note-taking app, CRM)

○ Result: A single noun with clear descriptors to qualify the minimum acceptable work result following the execution of all steps, such as "Confirmed product availability report (including quantity, timeframe, and any limitations)".

○ Capabilities: A bulleted list of skills, tools, or traits required for the fulfillment of the task, such as:
- Effective communication skills
- Ability to identify relevant production managers
- Knowledge of product lines and production processes
- Basic record-keeping skills

○ Completion criteria: A bulleted list of product specifications that determine whether an output meets minimum quality standards to declare the task completed, such as:
- Information clearly documented, including product, quantity, timeframe, and availability status.
- The product availability report accurately reflects current stock levels and expected replenishment dates.
- Confirmation from production managers is obtained and documented.

○ Completion instructions: A numbered list of activities guiding handoff of task products to relevant downstream or upstream stakeholders, such as:
1. Upload the updated product availability report to the shared drive.
2. Share confirmed product availability information with relevant stakeholders (e.g., sales team, inventory management).
3. File documentation for future reference.

○ Report Summary: A brief overview or summary, including the task name, the task product, and the task status, such as "Product Availability Confirmed: [Product Name], [Quantity Available], [Expected Replenishment Date], Status: Completed.

○ Report Language: Use the name of the language or dialect that is used to communicate about the task, such as *"English"*.

○ Report Medium: The name of the format or type of the task report, such as *"Text (Email and shared drive document)"*.

○ Delivery Mode: The name of the method or channel that is used to deliver the task report, such as *"Electronic, through Email, mail or internal system upload"*.

○ Status Options: A bulleted list of the possible states or conditions of the task, such as:
- *Completed.*
- *In progress.*
- *Canceled.*

- Step 5: Review the completed task specification template for any gaps, errors, or ambiguities, and revise it accordingly. It is often helpful to review this with another team member before using it to inform the selection of an AI agent or assigning the task to an agent (human or AI).

*Part 4: Identify or develop AI solutions*

The completed task specification template offers language and a checklist of necessary characteristics that are important when identifying, developing or adapting the right AI tool to integrate into an existing workflow. At this stage it is possible to use the information generated from the previous three parts to help the search for, to communicate needs to, or vet AI developers and their offerings.

**Automated method**

The automated procedure involves pasting pre-formulated prompts into available LLMs\*. For all automated options, it is advisable to test the prompt across various LLM options and compile or choose the most useful version, and then verify it with someone else.

*Part 1: Task (re)formulation*

The following text is an example of a prompt to generate a list of tasks. This should be modified depending on the LLM version, and the needs of the task:

> *I have a task description for my organization that I would like you to help me verify whether it meets ALL of the following criteria to be considered as the most basic unit of work:*
>
> - *Completable by a single agent independently (with or without consulting or others). An agent is any person or thing that can perform tasks. Within an organizational setting, every human is an agent, from members of the board through executives, to managers and front-line operators.*
> - *A single well-defined action represented by a single verb, without needing further breakdown.*
> - *A distinct identifiable result or output, either material or informational, which serves as a clear indicator of completion.*
>
> *Here is the initial task description: __________.*
>
> *Please follow these steps to verify whether this initial task description meets the criteria for a basic unit of work, or whether it needs to be broken down into and reformulated as single units of work:*
>
> - *Step 1: Use the criteria for the most basic unit of work to evaluate the task description, and check if it meets all the criteria. If yes, you are done. If not, go to the next step.*
> - *Step 2: Break down the task description into smaller and more manageable units of work, ensuring all the criteria for a basic unit of work are met, and generate a new list of tasks.*
> - *Step 3: Use the criteria for a basic unit of work to evaluate each unit, and check if it meets all the criteria. If not, repeat step 2 until all the units meet the criteria.*

- *Step 4: Once all of the units of work meet the criteria, include them in a final list labeled "Tasks". Write each unit of work in a clear and concise way, using the verb and the output as the main components.*

*Part 2: Assessing and ranking the ease of task automation*

Below is a generic prompt that managers, executives, or AI developers can use to assess the automation potential of a list of tasks using the Task Automation Index. This prompt is designed to be provided to a large language model like GPT-4/Bing/Copilot, Gemini/Bard or Claude for evaluation.

*Please assess the extent to which a given list of tasks within our organization are suitable for automation, using the Task Automation Index. This evaluation will help identify tasks that can be automated to improve efficiency, reduce errors, and optimize resource allocation.*

*List of Tasks: Below is a list of tasks for which we seek an automation potential assessment. Each task is briefly described to provide context on its nature and requirements.*

*Task 1: [Description]*

*Task 2: [Description]*

*Task 3: [Description]*

*(Add more tasks as needed)*

*Assessment criteria: For each task, please evaluate and score the following conditions based on the information provided and your understanding of typical automation capabilities:*

*Standardized input: Assess whether the input format and data types for the task are standardized. Score from 0 (not predefined) to 4 (completely standardized).*

*Well-defined rules: Determine if the rules for completing the task are clearly defined. Score from 0 (no rules) to 4 (explicitly defined and logically organized).*

*Repetitive: Evaluate the repetitiveness of the task. Score from 0 (unique) to 4 (always repetitive).*

*Data-dependent: Consider how much the task depends on data availability. Score from 0 (data not necessary) to 4 (data fully required and available).*

*Verifiable or measurable output: Judge if the task's output is verifiable or measurable. Score from 0 (not verifiable) to 4 (completely verifiable objectively).*

*Total score calculation: For each task, add up the scores from the five conditions to get a total score.*

*Ranking and recommendations: Based on the total scores, rank the tasks from most to least suitable for automation. Provide recommendations on which tasks are prime candidates for automation and which may require further analysis or modifications to enhance their automation potential.*

*Interpretation based on Task Automation Index total scores:*

*16-20: Highly suitable for automation*

*12-15: Suitable for automation*

*8-11: Moderately suitable for automation*

*4-7: Limited suitability for automation*

*0-3: Not suitable for automation*

*Part 3: Task specification template completion*

The following is the prompt designed to help complete the task specification template, based on ONLY ONE of the tasks generated with the previous prompt, or manually, using any of the publicly available LLMs*:

> *I would like you to help me complete a task specification template, which is a structure of data that provides detailed information in order to identify or adapt the best AI tools to successfully integrate them into existing workflows.*
>
> *Here is my task: "__________"*
>
> *Please complete the specification template, by specifying, assigning, and validating the 16 parts of the task, which are:*
>
> - *Unique ID: This is a value or code that identifies a specific completed task specification template, using numeric or alphanumeric strings.*
>
> - *Agent: The human or AI entity responsible for task completion (insert "TBD" if there is none assigned at the moment).*
>
> - *Deadline: This should be an unambiguous target calendar date, specified in MM/DD/YYYY numerics, representing the date before which the task should be completed*
>
> - *Action: The act is the action or behavior that is required to perform the task. It should be simple, concrete, and achievable. It should also be aligned with the task name, with its focus on one verb.*
>
> - *Steps: These are the instructions for sequential smaller activities that require orchestration to generate the main output. These should also be completable independently by a single agent, having a well-defined action represented by a single verb, and possessing a distinct, identifiable outcome, without needing further breakdown.*
>
> - *Frequency: Expected regularity of task activation based on the recurrence of the same task.*
>
> - *Materials: Checklist of raw work elements necessary for commencing and completing the task.*
>
> - *Result: A single noun with clear descriptors qualifying the minimum acceptable work result following the execution of all steps.*
>
> - *Capabilities: Enumeration of skills, tools or traits required for the fulfillment of the task.*
>
> - *Completion Criteria: Product specifications that determine whether an output meets minimum quality standards to declare the task completed.*
>
> - *Completion Instructions: Protocols guiding handoff of task products to relevant downstream or upstream stakeholders.*
>
> - *Report Summary: A brief statement that offers an overview or summary, including the task name, the task product, and the task status.*

- *Report Language: The report language is the language or dialect that is used to communicate about the task.*

- *Report Medium: The format or type of the task report, including text, audio, video or image.*

- *Delivery mode: The method or channel that is used to deliver the task report.*

- *Status Options: The possible states or conditions of the task, such as: completed, in progress, pending, canceled.*

*Note: The prompts were tested on GPT-4/Bing/Copilot, Claude and Bard/Gemini on January 31, 2024, and might need revision to match the development of the LLMs*

*Part 4: Identify or develop AI solutions*

Same as Part 4 in the manual section.

## Discussion

The accelerating pace of AI and robotics innovation is bringing about unprecedented convergence between management and computer science. The promises of hyper-productivity and growth are motivating profound changes in organizations across all industries and sectors that are reshaping the very essence of their operations.

Fulfilling the transformative potential of AI within organizational settings requires the navigation of complex multifaceted challenges, spanning the entire operational spectrum of organizations—from strategic planning at the executive level to the day-to-day activities of any enterprise—and extending through the entire AI development lifecycle—from initial concept through development, deployment, and ongoing management. We created Computational Management to address these challenges. By uniting the problem-solving techniques of computational thinking with the principles of management science, Computational Management seeks to revolutionize how workflows can be optimized with AI agents.

In an effort to be coherent with our work, we used LLMs as research assistants and consultants. We were very impressed by how much LLMs increased our capacity to search the scholarly and gray literature, by their ability to judge the comprehensiveness of the list of task specification template parts. Above all, we were surprised by the quality of their inputs and suggestions during the creation and refinement of the prompts to automate task (re)formulation, as well as to assess the potential for task automation and to complete the task specification template.

In the field of task automation with AI, our approach through task (re)formulation within the framework of Computational Management provides a dynamic, real-world application that goes beyond the static nature of traditional methods of task management, such as microtasking (29), task specification (30), task decomposition (31), or standard operating procedures (32). Our strategy not only acknowledges the strengths of these traditional methods, but also enhances them by incorporating the flexibility and learning capabilities of AI. In doing so, Computational Management bridges the gap between rigid task structuring and the fluid, evolving nature of AI, positioning it as a pivotal methodology for future AI-driven task automation efforts. By integrating task (re)formulation with the completion of the task specification template and employing the Task Automation Index, we offer a unique and practical approach to assessing and optimizing tasks for AI integration.

Computational Management also has limitations. First of all, the process and its related procedures are based on an initial theoretical conceptualization that requires empirical evaluation and validation in real-world settings, to substantiate their applicability, relevance and benefits across diverse use cases.

An additional limitation might result from the perception of managers or supervisors about the insufficient return on the additional effort required to embrace Computational Management, given the high rates of work overload and burnout they are already facing. To ease these concerns, it is essential to engage in a strategy focused on achieving quick wins and minimal disruption, through gradual application of the procedures to tasks with incremental levels of complexity.

Lastly, the incorporation of Computational Management into the day-to-day operations of organizations might be hindered by its close association with AI and the resistance that human agents might experience as a result of fears about being replaced by non-human intelligences (33).

## Future directions

As we venture into the uncharted intersection of organizational transformation and AI, there are several intriguing pathways that promise to redefine the landscape within traditional management practices.

One of these paths highlights the convergence between manual and cognitive tasks. Manual tasks, traditionally defined by their physical nature, have been the primary focus of automation through robotics and mechanical systems, aimed at improving efficiency, consistency, and safety in environments ranging from manufacturing floors to logistics operations. These tasks, characterized by repetitive, structured actions, lend themselves more readily to automation because they often involve clear, deterministic processes that can be explicitly programmed into machines. On the other hand, cognitive tasks, which encompass mental processes such as planning, decision-making, problem-solving, and creative thinking, present a different set of challenges and opportunities for automation. The development of LLMs has ushered in a new era where machines can not only process and analyze vast quantities of data at speeds unattainable by humans but also generate insights, make predictions, and even emulate creative processes to a certain degree. This capability to handle tasks that require understanding context, interpreting nuances, and generating coherent responses has expanded the scope of automation from purely manual tasks to domains that were once considered the exclusive domain of human intellect (34).

At this point in time where manual and cognitive tasks are blending, especially through robotics powered by multimodal AI, the range of possibilities for automation is expanding rapidly, raising essential questions about its desirability and impact. Automating manual tasks can certainly free humans from laborious, mundane, or dangerous work, while automating cognitive tasks stimulating deep reflection about the role of human judgment, the value of experience, and the importance of ethical considerations in decision-making processes. The rapid evolution of foundational AI models—particularly those that go beyond LLMs, such as large action models (LAM) or symbolic models—opens unprecedented opportunities to expand the scope of Computational Management from task automation to predictive analytics, decision support, and even strategic planning (35). This is making it more relevant than ever to strive for a balanced approach that respects the unique strengths of human cognition and the unparalleled efficiency and scalability of machines (36).

At the same time, another central theme emerges: the prospect of reverse engineering tasks traditionally carried out by humans. One of the main by-products of Computational Management is that it creates constructive pressure on organizational leaders to increase the precision, clarity and granularity of their ideas, intentions and expectations, with potential productivity gains that could be achieved even without introducing AI in their workplaces. This enhanced clarity could also support efforts to assess the value of AI solutions under optimal parity conditions, thanks to the allocation of tasks that are formulated in a truly executable way. This should be particularly relevant for any human tasks—and by default, any human jobs—that could be completed on a computer. In these cases, completed task specification templates could act as a particularly rich source of metadata for the algorithms that will guide the AI agents.

Because of its emphasis on first principles, Computational Management also paves the way to explore the evolutionary aspects of organizations, progressing from single agents completing single tasks within a specific organizational context, to multimodal hybrid (human-AI) swarms reshaping entire industries, sectors and economies, with the capability to pursue complex goals autonomously in highly-dynamic conditions (37). Along the way, these adaptive hybrid intelligence ecosystems could act as rich living laboratories for the development of new formal languages and ways of thinking for articulating the intentions, priorities, and values of human and AI agents, thus significantly enhancing clarity and alignment in organizational settings, and closing, even further, the gap between human intuition and algorithmic precision.

In sum, Computational Management emerges not just as a framework, but as a catalyst for a new era in which humans and AI agents could be deployed to make the most of their unique strengths in organizations. By focusing on bridging the gap between the boardroom and the codebase, between human intuition and algorithmic precision, Computational Management offers a roadmap and a new toolkit for navigating this transformative era, while contributing to forge a future in which humans and AI can thrive, together.

# References

1. Fernández-Macías E, Bisello M. A comprehensive taxonomy of tasks for assessing the impact of new technologies on work. Soc Indic Res. 2022 Jan;159(2):821–41.

2. Jadad AR, Arango A, Sepúlveda JD, Espinal S, Rodríguez D, Wind K, editors. Unleashing a pandemic of health from the workplace: Believing is seeing. Beati Inc; 2017. 235 p.

3. Acemoglu D, Restrepo P. Artificial intelligence, automation, and work. In: The economics of artificial intelligence: An agenda. University of Chicago Press; 2018. p. 197–236.

4. O*NET Online [Internet]. [cited 2024 Feb 4]. Sales Managers (11-2022.00). Available from: https://www.onetonline.org/link/summary/11-2022.00

5. Eberding LM, Belenchia M, Sheikhlar A, Thórisson KR. About the intricacy of tasks. In: Artificial General Intelligence. Cham: Springer International Publishing; 2022. p. 65–74.

6.  Jackendoff R. What is a Concept, that a Person May Grasp It? Mind Lang. 1989 Mar;4(1-2):68–102.

7.  Fonseca RJ, Weber GW, Telhada J, editors. Computational Management Science: State of the Art 2014 (Lecture Notes in Economics and Mathematical Systems, 682). 1st ed. Springer; 2015. 266 p.

8.  Patnaik S, Tajeddini K, Jain V, editors. Computational Management: Applications of Computational Intelligence in Business Management (Modeling and Optimization in Science and Technologies Book 18). 1st ed. Springer; 2021. 1147 p.

9.  Rustem B, Amman H, Maros I, Pardalos P. Computational Management Science. doc.ic.ac.uk.

10. Springer [Internet]. [cited 2023 Dec 4]. Computational Management Science. Available from: https://www.springer.com/journal/10287/aims-and-scope

11. Brogliato MS. Essays in computational management science. bibliotecadigital.fgv.br; 2018. Available from: https://sci-hub.st/https://bibliotecadigital.fgv.br/dspace/handle/10438/24615

12. Nanda P, Patnaik D, Patnaik S. Computational Management—An Overview. In: Patnaik S, Tajeddini K, Jain V, editors. Computational Management: Applications of Computational Intelligence in Business Management. Cham: Springer International Publishing; 2021. p. 3–21.

13. Lowe T, Brophy S. An operationalized model for defining computational thinking. In: 2017 IEEE Frontiers in Education Conference (FIE). IEEE; 2017. p. 1–8.

14. Wolfram S. Stephen Wolfram Writings. 2016. How to Teach Computational Thinking—Stephen Wolfram Writings. Available from: https://writings.stephenwolfram.com/2016/09/how-to-teach-computational-thinking/

15. Wing JM. Computational thinking. Commun ACM. 2006 Mar;49(3):33–5.

16. Lodi M, Martini S. Computational Thinking, Between Papert and Wing. Science & Education. 2021 Aug 1;30(4):883–908.

17. O*NET OnLine. Job Duties Custom List. Available from: https://www.onetonline.org/help/online/task

18. O*NET® in Action. Available from: https://www.onetcenter.org/action.html

19. World Economic Forum. The Future of Jobs Report 2023. Available from: https://www.weforum.org/publications/the-future-of-jobs-report-2023/

20. Kochhar R. Pew Research Center. 2023. Methodology for O*NET analysis. Available from: https://www.pewresearch.org/social-trends/2023/07/26/2023-ai-and-jobs-methodology-for-onet-analysis/

21. Huang Q, Shen Y, Sun Y, Zhang T. The Layoff Generation: How Generative Ai Will Reshape Employment and Labor Markets. 2023. Available from: https://papers.ssrn.com/abstract=4534294

22. Felten E, Raj M, Seamans R. How will language modelers like ChatGPT affect occupations and industries?. arXiv [econ.GN]. 2023. Available from: http://arxiv.org/abs/2303.01157

23. Zhang S, Nickerson JV. The Coevolution of Tasks and Technologies. In: Academy of

Management Annual Meeting. par.nsf.gov; 2022. Available from: https://sci-hub.st/https://par.nsf.gov/biblio/10334734

24. Rodrigues M, Fernández-Macías E, Sostero M. A unified conceptual framework of tasks, skills and competences. econstor.eu; 2021. Available from: https://sci-hub.st/https://www.econstor.eu/handle/10419/231348

25. Autor DH. The "Task Approach" to Labor Markets: An Overview. National Bureau of Economic Research; 2013 Jan. (Working Paper Series). Report No.: Discussion Paper No. 7178. Available from: http://www.nber.org/papers/w18711

26. Wood RE. Task complexity: Definition of the construct. Organ Behav Hum Decis Process. 1986 Feb 1;37(1):60–82.

27. Fernández-Macías E, Bisello M. Measuring the content and methods of work: a comprehensive task framework. {European Foundation for the Improvement of Living and Working Conditions}; 2017. Available from: https://sci-hub.st/https://ibs.org.pl/app/uploads/2017/10/D1.-Martina-Bisello.pdf

28. O*NET Online. Public Relations Specialists (27-3031.00). Available from: https://www.onetonline.org/link/summary/27-3031.00

29. Zulfiqar M, Malik MN, Khan HH. Microtasking activities in crowdsourced software development: A systematic literature review. IEEE Access. 2022;10:24721–37.

30. Agrawal P. The task specification problem. In: Conference on Robot Learning. PMLR; 2022. p. 1745–51.

31. Stedmon AW. Human factors methods: a practical guide for engineering and design (second edition). Ergonomics. 2014 Nov 2;57(11):1767–9.

32. Akyar I. Standard Operating Procedures (What Are They Good For ?). Latest Research into Quality Control. InTech; 2012. Available from: http://dx.doi.org/10.5772/50439

33. Van Quaquebeke N, Gerpott FH. The now, new, and next of digital leadership: How Artificial Intelligence (AI) will take over and change leadership as we know it. J Leadersh Organ Stud. 2023 Aug;30(3):265–75.

34. Piton C. The economic consequences of artificial intelligence: an overview. NBB economic review. 2023;1–29.

35. Schneider J, Meske C, Kuss P. Foundation Models: A New Paradigm for Artificial Intelligence. BISE. 2024 Jan 29:1-1.

36. Pasquinelli M. The Eye of the Master: A Social History of Artificial Intelligence. Verso Books; 2023.

37. Wang GY, Cheng DD, Xia DY, Jiang HH. Swarm Intelligence Research: From Bio-inspired Single-population Swarm Intelligence to Human-machine Hybrid Swarm Intelligence. Mach Intel Res. 2023 Feb 1;20(1):121–44.

# Appendix A. Search strategies

| Target | Search string |
|---|---|
| Existing knowledge, of any kind, on 'computational management' | *intitle:"computational management"* |
| Conceptual frameworks focused on 'computational thinking' | *intitle:"conceptual framework" AND intitle:"computational thinking"* |
| Conceptual frameworks focused on 'management science' | *(intitle:"management science" OR intitle:"business management" OR intitle:"organizational management" OR intitle:"human resource management" OR intitle:"workplace management" OR intitle:"corporate management" OR intitle:"scientific management") AND intitle:"conceptual framework"* |
| Conceptual frameworks focused on tasks | *(intitle:task OR intitle:tasks) AND (intitle:"conceptual framework" OR intitle:"conceptual frameworks")* |

# Appendix B. Original task descriptions for corporate communication specialists listed on O*NET (28)

Respond to requests for information from the media or designate an appropriate spokesperson or information source.

Plan or direct development or communication of programs to maintain favorable public or stockholder perceptions of an organization's accomplishments, agenda, or environmental responsibility.

Post and update content on the company's Web site and social media outlets.

Write press releases or other media communications to promote clients.

Establish or maintain cooperative relationships with representatives of community, consumer, employee, or public interest groups.

Confer with other managers to identify trends or key group interests or concerns or to provide advice on business decisions.

Coach client representatives in effective communication with the public or with employees.

Study the objectives, promotional policies, or needs of organizations to develop public relations strategies that will influence public opinion or promote ideas, products, or services.

Prepare or edit organizational publications, such as employee newsletters or stockholders' reports, for internal or external audiences.

Arrange public appearances, lectures, contests, or exhibits for clients to increase product or service awareness or to promote goodwill.

Plan or conduct market or public opinion research to test products or determine potential for product success, communicating results to client or management.

Develop plans or materials to communicate organizational activities that are beneficial to the environment, public safety, or other important social issues.

Confer with production or support personnel to produce or coordinate production of advertisements or promotions.

Consult with advertising agencies or staff to arrange promotional campaigns in all types of media for products, organizations, or individuals.

Prepare or deliver speeches to further public relations objectives.

Coordinate public responses to environmental management incidents or conflicts.

Develop marketing campaigns for environmental technologies or services.

Purchase advertising space or time as required to promote client's product or agenda.

## Appendix C. Task Automation Index

Use this Task Automation Index to generate a total score representing the extent to which a task reflects five conditions for automation. Each condition has a set of statements about a task with a designated amount of points for each.

1. Create a table with seven columns and the amount of rows that represents the number of tasks that you are evaluating, plus an additional row for the headers.

2. In the top row of the table, label the columns with the following headings from left to right: "Task", "Standardized input", "Well-defined rules", "Repetitive", "Data-dependent", "Verifiable or measurable output" and "Total score".

3. Fill out the first column on the left, under the "Task" heading, with the list of tasks you are evaluating.

4. Beginning from the top of the list, select the statement below that best describes the the extent to which the task reflects the specific condition for automation, and write down the number of points under the corresponding heading:

    - Standardized Input: Select the statement that best describes the extent to which the agent that will complete the task receives information that is consistent, structured, and presented in a format that can be accurately understood and processed, and note the corresponding points:

        | | |
        |---|---|
        | 0 points: | Input format is not predefined and data types are completely inconsistent. |
        | 1 point: | Input format and data types are mostly inconsistent. |
        | 2 points: | Input format and data types are somewhat standardized. |
        | 3 points: | Input format is mostly predefined and data types are consistent with rare exceptions. |
        | 4 points: | Input format is predefined specifying the order and naming of fields with consistent data types |

    - Well-Defined Rules: Select the statement that best describes the extent to which the steps to complete a task are unambiguous and organized in a logical order, and note the corresponding points:

        | | |
        |---|---|
        | 0 points: | There are no rules to complete the task. |
        | 1 point: | The rules are ambiguous. |
        | 2 points: | Rules are explicit for most scenarios, with areas of ambiguity. |
        | 3 points: | Rules are explicit, with very few ambiguous aspects, and are mostly organized in a logical order. |
        | 4 points: | Rules are explicitly defined and organized in a logical order. |

    - Repetitive: Select the statement that best describes the extent to which the task is performed frequently, involving similar processes or actions each time, and note the corresponding points:

0 points: Task is unique or never repeated.
1 point: Task is occasionally repeated.
2 points: Task is often repeated.
3 points: Task is frequently repeated.
4 points: Task is always repetitive.

- Data-Dependent: Select the statement that best describes the extent to which data availability is required to complete the task, and note the corresponding points:

    0 points: Data and not necessary for task completion.
    1 point: Data somewhat required or unavailable for task completion.
    2 points: Data required and somewhat available for task completion.
    3 points: Data required and mostly available for task completion.
    4 points: Data required and fully available for task completion.

- Objective Output: Select the statement that best describes the extent to which the task's result can be verified, and note the corresponding points:

    0 points: Output does not have a metric or benchmark, or cannot be verified
    1 point: Output has limited metrics or benchmarks, or verification is mainly subjective
    2 points: Output has clear metrics or benchmarks, but verification is mainly subjective.
    3 points: Output has clear metrics or benchmarks and is mostly verifiable objectively.
    4 points: Output has a clear metric or benchmark and is completely verifiable objectively.

5. Add up the total points in each row and write the result for each in the last column, under the heading "Total score".
6. In a list, or in the same table, rearrange the tasks in order of most automatable (those with the higher score) to least (those with the lower scores).
7. Interpret the results using the following suggested guidelines, included The Task Automation Index:

Suggested total score interpretation (range: 0 to 20 points)

16-20: Highly Suitable for Automation

Tasks scoring in this range are excellent candidates for automation, indicating a high degree of standardization, clear rules, repetitiveness, data-dependency, and measurable outputs.

12-15: Suitable for Automation

These tasks are good candidates for automation, showing strong potential with some areas that might benefit from minor adjustments or refinements.

8-11: Moderately Suitable for Automation

> Tasks within this score range are possible to automate but might require significant modifications to enhance their automation potential or might be more complex to automate effectively.

4-7: Limited (Low?) Suitability for Automation

> Such tasks have limited automation potential due to significant variability, ambiguity, low repetition, or subjective outputs. Automation may be feasible for parts of the task or in a limited capacity.

0-3: Not Suitable for Automation

> Tasks scoring in this lowest range are currently not viable for automation with existing technologies, likely requiring a high degree of human judgment, creativity, or adaptability.